\documentclass[aps,pra,twocolumn,amsmath,amssymb]{revtex4}
\usepackage{graphicx}
\usepackage{amsmath}
\usepackage{bm}
\usepackage{times}
\usepackage{dcolumn}
\usepackage{bm}
\newcommand{\be}{\begin{equation}}
\newcommand{\ee}{\end{equation}}
\newcommand{\ba}{\begin{array}}
\newcommand{\ea}{\end{array}}
\newcommand{\bea}{\begin{eqnarray}}
\newcommand{\eea}{\end{eqnarray}}

\def\openone{\leavevmode\hbox{\small1\kern-3.8pt\normalsize1}}%

\begin{document}

\title{General framework for estimating the ultimate precision limit in noisy quantum-enhanced metrology}
\author{B. M. Escher}\email{bmescher@if.ufrj.br}\author{R. L. de Matos Filho}\author{L. Davidovich}

\affiliation{Instituto de F\'{\i}sica, Universidade Federal do Rio de Janeiro, 21.941-972, Rio de Janeiro (RJ) Brazil}

\date{\today}

\begin{abstract}
The estimation of parameters characterizing dynamical processes is central to science and technology. The estimation error decreases with the number $N$ of resources employed in the experiment (which could quantify, for instance, the number of probes or the probing energy). Typically, it scales as $1/\sqrt{N}$. Quantum strategies may improve the precision, for noiseless processes, by an extra factor $1/\sqrt{N}$. For noisy processes, it is not known in general if and when this improvement can be achieved. Here we propose a general framework for obtaining attainable and useful lower bounds for the ultimate limit of precision in noisy systems. We apply this bound to lossy optical interferometry and atomic spectroscopy in the presence of dephasing, showing that it captures the main features of the transition from the $1/N$ to the $1/\sqrt{N}$ behavior as $N$ increases, independently of the initial state of the probes, and even with use of adaptive feedback.
\end{abstract}

\maketitle

It is by now well established that, in the absence of noise, quantum properties of the probes help to increase the precision in the estimation of parameters that characterize dynamical processes, like phase shifts in optical interferometry or transition frequencies in atomic spectroscopy \cite{Bollingerpra96,Dowlingjmodopt02,Giosci04, Gioprl06}. The estimation error decreases with the amount of resources  employed in the measurement, which might be for instance the energy of a probing light field or the number $N$ of identical probes. For independent probes,  it is proportional to $1/\sqrt{N}$, a consequence of the central-limit theorem. By entangling the probes one may attain, for noiseless processes, the ultimate lower bound for the estimation error, which scales then with $1/N$\cite{Giosci04, Gioprl06}, the so-called Heisenberg limit, thus allowing better accuracy for the same number of resources. For noisy processes, however, it is not known in general if this limit can be attained, and if entanglement can still be a helpful resource for this purpose. For lossy optical interferometry, it was shown recently \cite{jan, sergey} that the Heisenberg limit is not attainable when $N\rightarrow\infty$. General expressions for the uncertainty in the estimation are known, but their calculation involves complex optimization procedures, which become quite cumbersome when the number  of resources increases. Here we show that the effectiveness of quantum states for parameter estimation in the presence of noise can be precisely assessed. We introduce a bound for this uncertainty, proven to be attainable, which leads to useful expressions for the ultimate limit of precision in noisy systems. We exemplify the utility of this bound by applying it to lossy optical interferometry and atomic spectroscopy in the presence of dephasing.

\begin{figure}[t]
\centering
\includegraphics[width=0.50\textwidth]{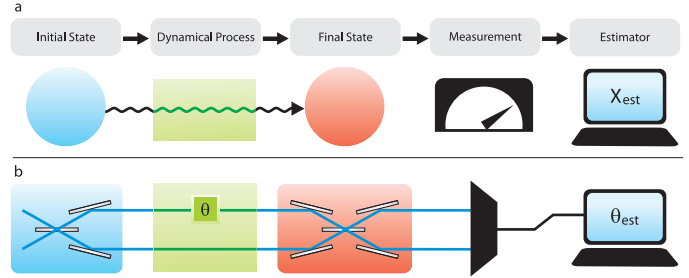}
\caption{{\bf Setups for quantum parameter estimation.} a) General algorithm to estimate an unknown parameter $x$ of an arbitrary dynamical process. The probe, prepared in a known initial state, is sent through a physical channel. A measurement is performed on the final state, from which the parameter $x$ is estimated. b) Setup for estimating a phase shift $\theta$ in an optical interferometer.}
\label{erptsQFIt}
\end{figure}
 
\noindent {\bf Parameter estimation and the Cram\'er-Rao bound}\\
A general protocol to estimate an unknown parameter $x$ corresponding to a quantum process is shown in Fig.~1a). This parameter is estimated from the knowledge of the initial and final states of a probe that undergoes the process under investigation. The protocol may be divided in three stages. First, the probe is prepared in an initial state (first box on the left-hand side of the figure), and evolves under the action of the quantum process, corresponding to the second box in Fig.~1a). In general, after this step, the probe will be in a mixed state (third box). The second stage concerns the choice of a suitable measurement, applied to the probe after its evolution (fourth box), so as to extract information about the parameter to be estimated. The third and last step consists in associating, through some rule (estimator), each experimental result with an estimation of the parameter.  A concrete example is shown in Fig.~1b), which refers to phase shift estimation in optical interferometry.

Typically,  one has some prior knowledge about the parameter. This is sometimes called in the literature ``local estimation'', as opposed to the situation when there is complete ignorance about the parameter to be estimated, which is known as ``global estimation.'' In local estimation, a merit quantifier that allows the comparison of different strategies and performances is the error estimate \cite{Helstrombook76,Holbook82} $\delta x \equiv \sqrt{\langle \left( x_{\rm est} - x_{\rm real} \right)^{2} \rangle}$, where the average is taken over all possible experimental results, $x_{\rm real}$ is the real value of the unknown parameter $x$ and $x_{\rm est}$ is the estimated value of $x$, obtained from the measurement results through the use of an estimator.

When the initial state, the physical system, and the measurement procedure, involving $\nu$ repetitions of the experiment, are fixed, the error $\delta x$ is limited by the Cram\'{e}r - Rao inequality \cite{Cramerbook46} $\delta x \geq 1 / \sqrt{\nu F(x_{real})}$ (valid for unbiased estimators, that is, estimators for which $\langle x_{\rm est}\rangle=x_{\rm real}$).  Here $F(x)$ is the Fisher Information, given by
\begin{equation}\label{fisher}
F(x)=\sum_j p_j(x)\left\{\frac{d\ln\left[p_j(x)\right]}{dx}\right\}^2\,,
\end{equation}
where $p_j(x)$ is the probability of getting the experimental result $j$. This relation is valid for both classical and quantum physics.

In quantum mechanics, $p_j(x)={\rm Tr}\left[\hat\rho(x)\hat E_j\right]$, where the operators $\hat E_j$ correspond to a specific measurement setup, associated to the results $j$, and $\hat\rho(x)$ is the density matrix of the probe after its interaction with the system under investigation. The Hermitian operators $\hat{E}_j$ are positive operator-valued measures (POVM's) \cite{Nielsenbook01}, satisfying the relation $\sum_j \hat E_j=\openone$. In general, the evolution of a density matrix $\hat\rho$ under the action of a quantum channel may be expressed in terms of Kraus operators \cite{Krausbook83} $\hat\Pi_\ell(x)$ as $\hat\rho(x)=\sum_\ell\hat\Pi_\ell(x)\hat\rho_0\hat\Pi_\ell^\dagger(x)$, where $\hat\rho_0$ is the initial value of $\hat\rho$ and $\sum_{\ell}\hat\Pi^{\dagger}_{\ell}(x)\hat\Pi_{\ell}(x)=\openone$. The maximization of (\ref{fisher}) over all possible measurement strategies  yields the so-called Quantum Fisher Information ${\cal F}_{Q}$ \cite{Helstrombook76,Holbook82,Braunsprl94,Braunsteinannals96} and the quantum generalization of the Cram\'er-Rao inequality,
\begin{equation}\label{deltaxfq}
\delta x\ge 1/\sqrt{\nu {\cal F}_Q\left[\hat\rho\left(x_{\rm real}\right)\right]}\,.
\end{equation}

For a closed system, evolving under a unitary transformation $\hat{U}(x)$ and prepared in a pure initial state $\hat{\rho}_{0}=\vert\psi\rangle\langle\psi\vert$, ${\cal F}_{Q}$ can be expressed as \cite{Boixoprl07}
\begin{equation}\label{equni}
{\cal F}_{Q}\left[\hat{U}(x)\hat{\rho}_{0}\hat{U}^\dagger(x) \right] =4 \langle\Delta\hat H^2\rangle\ ,
\end{equation}
where
\begin{equation*}
 \langle\Delta\hat H^2\rangle\equiv\left[\langle\psi\vert\hat{H}^{2}(x)\vert\psi\rangle - \langle\psi\vert\hat{H}(x)\vert\psi\rangle^2\right] ,
\end{equation*}
and $\hat{H}(x)\equiv i \left(d \hat{U}^{\dagger}(x) / dx \right) \hat{U}(x)$. However for more general situations where the initial state is a mixture and/or the evolution is not governed by a unitary transformation, a closed expression for ${\cal F}_{Q}$ as a function of Kraus operators is unknown and finding it remains an open problem.

Although analytic expressions for ${\cal F}_{Q}$ have been found for some specific classes of initial states and  non-unitary processes (see, for instance refs.~\cite{Monrasprl07,Dornerprl09,Dempra09}), in most cases only an upper bound to ${\cal F}_{Q}$ can be obtained. For example, when the initial state of the probe is mixed but the quantum channel is unitary, an upper bound based on  the convexity  of ${\cal F}_{Q}$ can always be established.  More recently  an  upper bound  to   ${\cal F}_{Q}$,  applicable to general quantum channels,  was derived in terms of  Kraus operators  and was shown to be attainable for a special class of processes (``quasi-classical" processes)~\cite{Sarovarjpmga06} .

\noindent {\bf Bounds for error estimation in noisy systems}\\
We introduce here an universal upper bound to ${\cal F}_{Q}$, expressed in terms of the initial state of the probe and any Kraus representation of the quantum channel. This bound is valid for both unitary and non-unitary processes, and it is always possible to choose a Kraus representation such that it coincides with the quantum Fisher information. We show that this bound leads to useful analytical approximations for the precision of parameter estimation.

Let $S$ be the probe used to estimate a parameter of a general dynamical process. Our strategy consists in introducing additional degrees of freedom, which play the role of an environment $E$ for the system $S$, so that a general dynamical evolution concerning $S$ is transformed into a unitary evolution for $S+E$. The problem is thus reduced to parameter estimation for a unitary evolution. A given general dynamical evolution of a system $S$ may be related to an infinitude of unitary evolutions of an enlarged system consisting of $S$ plus some ``environment" $E$. Fixing the unitary evolution and the ``environment" $E$ is equivalent to singling  a specific representation of  Kraus operators out from the unlimited number of such representations that describe the same dynamical evolution of $S$ alone. 

The detailed calculation of this upper bound can be found in the Supplementary Material. An outline of the derivation is presented in the Methods section. Given an initial pure state $\hat{\rho}_{0}=\vert\psi\rangle\langle\psi\vert$ of a system $S$ and an arbitrary process, which changes the state to $\hat{\rho}(x)\equiv\sum_{\ell}\hat{\Pi}_{\ell}(x)\hat{\rho}_{0}\hat{\Pi}^{\dagger}_{\ell}(x)$, we get the upper bound to ${\cal F}_Q$
\begin{equation}
C_{Q}\left(\hat{\rho}_{0}, \hat{\Pi}_{\ell}\left(x \right) \right) = 4 \left[ \langle \hat{H}_{1}(x) \rangle - \langle \hat{H}_{2}(x) \rangle^{2} \right] ,
\label{eqtot}
\end{equation}
where
\begin{equation}\label{h1}
 \hat{H}_{1}(x) \equiv \sum_{\ell} \frac{d \hat{\Pi}_{\ell}^{\dagger}\left(x \right)}{d x} \frac{d \hat{\Pi}_{\ell}\left(x \right)}{d x} ,
\end{equation}
\begin{equation}\label{h2}
 \hat{H}_{2}(x) \equiv i \sum_{\ell} \frac{d \hat{\Pi}_{\ell}^{\dagger}\left(x \right)}{d x} \hat{\Pi}_{\ell}\left(x \right) ,
\end{equation}
the symbol $\langle\phantom{\bullet}\rangle$ meaning $\langle\bullet\rangle\equiv Tr \left[ \bullet \hat{\rho}_{0} \right]$.

Equations (\ref{eqtot})-(\ref{h2}) show that the upper bound $C_Q$ can be explicitly evaluated in terms of the Kraus operators describing the quantum channel and the initial state of the probe.  When the process is unitary $C_Q$ reduces to the quantum Fisher information, as given by Eq.~(\ref{equni}). This upper bound can also be applied to situations with ancillas, where the initial state of the probe is entangled with some external system. The bound is then easily calculated from the reduced density matrix of the probe. The attainability of this bound is demonstrated in the Methods section, implying that 
\begin{equation}\label{fqcq}
{\cal F}_Q\left[\hat\rho(x)\right]={\rm min}_{\{\hat\Pi_\ell(x)\}}C_Q\left[\hat\rho_{0},\hat\Pi_\ell(x)\right]\,,
\end{equation}
where the minimization runs over all Kraus representations $\{\hat\Pi_\ell(x)\}$ of the quantum channel. 

Equation (\ref{fqcq}) leads to an interesting physical insight, within the framework of our method: there is always an environment $E$ such that monitoring it together with the system $S$ does not lead to more information about the parameter $x$ than that obtained by monitoring just the system $S$ itself. These considerations motivate a strategy for choosing convenient Kraus operators: the aim should be to reduce the non-redundant information about the parameter in the environment. 

We apply now the bound (\ref{eqtot}) to some important problems in quantum metrology, and show that it leads to the elucidation of fundamental questions regarding limits of precision in parameter estimation. 

\noindent {\bf Role of entanglement in quantum-enhanced metrology}\\
In recent years, much effort has been made to establish the relevance of entanglement in quantum metrology \cite{Bollingerpra96,Giosci04,Gioprl06,Dowlingjmodopt02,Ciracprl97,dowling,higgins,caves07,dowling2,Dornerprl09,Dempra09,bana,kac}. For unitary processes, it is now clear that entanglement does lead to enhancement of the precision in parameter estimation. However, an important question remains: does this gain persist in the presence of decoherence? 

A typical paradigm of quantum metrology involves  $N$ probes undergoing $N$ independent and identical $x$-dependent processes, and being  submitted to a measurement afterwards, with the aim of estimating the parameter $x$. If the initial state of the probes is not entangled and each probe is measured independently of the others, then the error scales at most as $\delta x \sim 1 / \sqrt{N}$ \cite{Giosci04}. It was shown in \cite{Gioprl06} that, for unitary processes and in the absence of feedback, initial entanglement is a necessary ingredient to improve the error scale to the ultimate quantum limit $\delta x \sim 1 / N$. Moreover, it was also shown that, for unitary processes of the type $\hat{U}(x)=\exp{ \left(i x \hat{H} \right)}$, it is always possible to find an initial entangled state of the probes that leads to $\delta x \sim 1 / N$.  

We consider now the effect of noise on these results.   When the $N$ probes are submitted to $N$ independent processes, which depend on the parameter $x$, a set of Kraus operators that describes the evolution of the probes is
\begin{equation}\label{product}
 \hat{\Pi}_{\ell_{1},\ell_{2},\ldots,\ell_{N}} (x) = \hat{\Pi}_{\ell_{1}}^{(1)}(x)\otimes\hat{\Pi}_{\ell_{2}}^{(2)}(x)\otimes\cdots\otimes\hat{\Pi}_{\ell_{N}}^{(N)}(x) ,
\end{equation}
where $\hat{\Pi}_{\ell_{m}}^{(m)}(x)$ is a Kraus operator corresponding to the $m$-th process. From (\ref{eqtot}), it is then straightforward to show that $C_{Q}$ can be decomposed into two parts (see Supplementary Material)
\begin{eqnarray} \label{resm} 
C_{Q}\left(\hat{\rho}_{0},\hat{\Pi}_{\ell_{1},\ell_{2},\ldots,\ell_{N}} (x)  \right) = 4 \sum_{m=1}^{N}\left[ \langle \hat{H}_{1}^{(m)} \rangle - \langle \hat{H}_{2}^{(m)} \rangle^{2} \right]\nonumber \\ 
 + 8 \sum_{m_{1}=2}^{N} \sum_{m_{2}=1}^{m_{1}-1} \left[\langle \hat{H}_{2}^{(m_{1})}\hat{H}_{2}^{(m_{2})} \rangle - \langle \hat{H}_{2}^{(m_{1})} \rangle\langle \hat{H}_{2}^{(m_{2})} \rangle\right], 
\end{eqnarray}
where the operators $\hat{H}_{1,2}^{(m)}$ correspond to the definitions (\ref{h1}) and (\ref{h2}) for the $m$-th process. The first  term in Eq. (\ref{resm}) is of $o(N)$, while the second is of $o(N^2)$, as they involve the sum of $N$ and $N(N-1)/2$ terms respectively. 

Since $C_Q\ge{\cal F}_Q$, this result has two immediate implications: (i) a necessary condition for $\delta x\sim 1/N$ is that the second line of (\ref{resm}) is different from zero, which establishes the kind of correlation needed in the initial state of the $N$ probes in order that the Heisenberg limit is attained -- this condition is also sufficient if there is a Kraus representation in the form (\ref{product}) that minimizes $C_Q$; (ii) if  a quantum channel has some Kraus representation for which  $\hat{H}_{2}^{(m)}=0$, then $\delta x\sim 1/\sqrt{N}$ at most, even when the initial state is entangled. The fact that $\hat{H}_{2}^{(m)}=0$ implies $\delta x\sim 1/\sqrt{N}$ was derived  for the more restrict situation of finite-dimensional spaces in \cite{Fujiwarajpmt08}, which also showed that the condition  $\hat{H}_{2}^{(m)}=0$ holds for almost all finite-dimensional quantum channels, the ``full-rank''  channels.  Moreover, (ii) remains valid when feedback is included, as demonstrated in the Supplementary Material. Therefore, for this class of channels, the limit $\delta x\sim1/N$ cannot be achieved, in the presence of decoherence, even with the use of entanglement and/or feedback control. 

\noindent {\bf Precision limits for lossy optical interferometry}\\
Optical interferometry with standard light sources leads to an uncertainty in the determination of the phase  that scales with the inverse of the square root of the mean number of photons, the so-called shot-noise or standard quantum limit \cite{Carltonprl80,Carltonprd81}. On the other hand,  it has been shown that squeezed \cite{Carltonprl80,Carltonprd81} or entangled states \cite{Giosci04,Dowlingjmodopt02} may lead, in the absence of losses, to a scaling of this uncertainty with the inverse of the mean number of photons. This is the ultimate limit imposed by quantum mechanics \cite{Gioprl06,Ou}. 

In the presence of losses, general results concerning the limit of precision are still unknown. Special entangled quantum states have been proposed for optical metrology in order to mitigate the deleterious effect of noise on the precision of phase estimation \cite{dowling2}. Numerical limits for the minimum uncertainty in the measurement of the phase shift, when either one or both arms of an interferometer are subject to photon losses, were found in \cite{Dornerprl09,Dempra09} for input states with fixed photon numbers. These studies show that, for states with total number of photons up to $N=80$, the best possible scaling of the uncertainty is intermediate between the standard and the Heisenberg limit. Experimental confirmation with two-photon entangled states was presented in \cite{kac}, while bounds found in \cite{jan} for global estimation and in \cite{sergey} for local estimation lead to a $1/\sqrt{N}$ scaling.  This raises an important question concerning the behavior of the precision as a function of $N$.Here we answer this question in a general way, by finding  an analytical lower bound for the phase uncertainty. 

We consider for definiteness a two-arm interferometer with a dispersive object placed in the upper arm (see Fig.~\ref{erptsQFIt}-b). An incoming photon is described by a two-mode state, each mode corresponding to one of the arms of the interferometer. In the absence of losses the initial two-mode state in the interferometer evolves into $|\psi(\theta)\rangle=\hat U(\theta)|\psi\rangle_{0}=\exp(i\hat n\theta)|\psi\rangle_{0}$, where $\hat n$ is the photon number operator corresponding to the dispersive-arm mode and $\theta$ is the phase shift parameter to be estimated. Therefore, from (\ref{equni}), 
\begin{equation}\label{lossless}
{\cal F}_Q\left(\hat{\rho}_{0}\right)=4\langle\Delta\hat n^2\rangle_0 \quad {\rm and}\quad \delta\theta\ge\left(4\nu\langle\Delta\hat n^2\rangle_0\right)^{-1/2} \ ,
\end{equation}
where $\langle\Delta\hat n^2\rangle_0$ is the photon-number variance of the state in the dispersive  arm of the interferometer, before it undergoes the phase shift. From this expression, it follows that states maximizing the photon-number variance lead to minimum values of the phase uncertainty. This is the case of the NOON states $|\psi(N)\rangle=(|N,0\rangle+|0,N\rangle)/\sqrt{2}$ \cite{Dowlingjmodopt02,dowling}, where $|N_1,N_2\rangle$ represents a Fock state with $N_1$ and $N_2$ photons in the upper and lower arms, respectively. For this state, $\langle\Delta\hat n^2\rangle_{\rm NOON}=N^2/4$, which yields the Heisenberg limit $\delta\theta\ge1/(\sqrt{\nu} N)$.

\begin{figure}[t]
\centering
\includegraphics[width=0.50\textwidth]{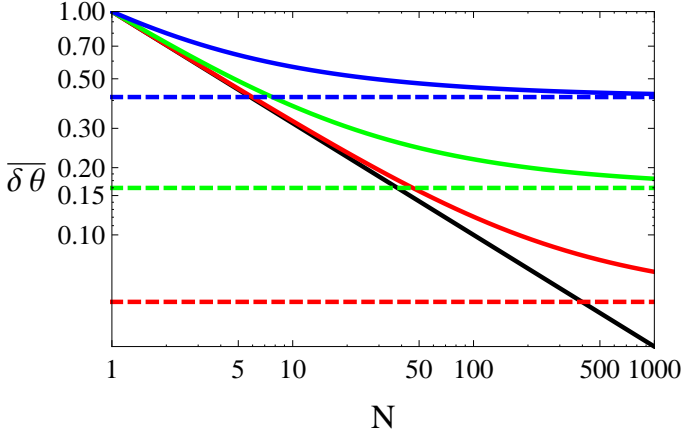}
\caption{{\bf Lower bounds for the phase error.} Lower bound $\overline{\delta\theta}$ of the normalized variance $\delta\theta\frac{\sqrt{4 \nu\eta N}}{1+\sqrt{\eta}}$ as a function of $N$ for $\eta=0.5$ (blue), $\eta=0.9$ (green), $\eta=0.99$ (red), and $\eta = 1$ (black). The respective dashed lines correspond to the limit (\ref{dtheta}) for each value of $\eta$.}
\label{main}
\end{figure}

We consider here interferometers with losses only in the dispersive arm  -- see  Fig.~1b). Generalization to losses in both arms is presented in the Supplementary Material. As shown in the Methods section, proper  choice of Kraus operators, inspired by physical considerations, leads to
\begin{equation}
{\cal F}_{Q} (\hat{\rho}_{0}) \leq  \left[ \frac{4 \eta \langle \hat{n} \rangle_0\langle \Delta \hat{n}^2 \rangle_0}{\langle \Delta \hat{n}^2 \rangle_0 \left(1 - \eta\right) + \eta\langle \hat{n} \rangle_0}\right] \,,
\label{main1}
\end{equation} 
where $\eta$ quantifies the photon losses (from $\eta=1$, lossless case, to $\eta=0$, complete absorption) and $\langle\hat n\rangle_0$ is the initial average number of photons in the dispersive arm.

The above equation has a high degree of generality: it was derived under no assumption whatsoever about the initial state, which might therefore have a definite, limited or unlimited number of photons. Furthermore,  it displays the limit of precision as a continuous function of losses as well as the variance and the average photon number of the field in the dispersive arm, for the entire range of values of these quantities. When $\langle\Delta\hat{n}^2 \rangle_{0} / \langle\hat{n} \rangle_{0} << \eta / (1-\eta)$, this expression approaches the lossless case~(\ref{lossless}),while when $\langle\Delta\hat{n}^2 \rangle_{0} / \langle\hat{n} \rangle_{0} >> \eta / (1-\eta)$ Eq.~(\ref{main1}) yields the more restricted result
\begin{equation}
\delta\theta\ge\sqrt{\frac{1-\eta}{4\nu\eta\langle\hat{n}\rangle_0}} \ .
\label{dtheta}
\end{equation}
Equations (\ref{main1}) and (\ref{dtheta})  imply, quite generally,  that photon losses in interferometers gradually blur the gain yielded by the special quantum states for phase estimation; even with the best strategy, asymptotically the improvement with respect to standard light sources is not by a scale change but only by a limited constant factor, as shown in (\ref{dtheta}). This last expression coincides with the bounds obtained in \cite{jan,sergey} for states with definite photon number. 

\begin{figure}[t]
\centering
\includegraphics[width=0.50\textwidth]{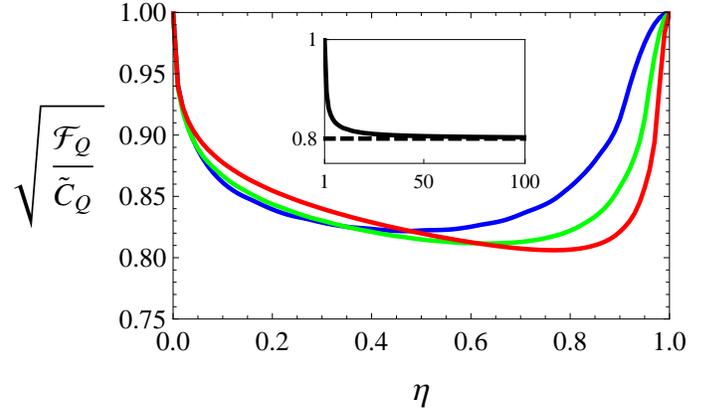}
\caption{{\bf Numerical test of the tightness of the bound.} Comparison between the numerical maximum value of ${\cal F}_{Q}$ and the upper bound $\tilde C_{Q}$, $\sqrt{{\cal F}_{Q} / \tilde{C}_{Q}}$ as a function of $\eta$, for $N=10$ (blue), $N=20$ (green), and $N=40$ (red). The inset displays the minimum of this ratio over all values of $\eta$ as a function of $N$.}
\label{fig3}
\end{figure}

For states with well-defined number $N$ of photons, extensively discussed in  quantum metrology, Eq.~(\ref{main1}) leads to an upper bound for the quantum Fisher information maximized over all initial states, which depends only on $\eta$ and on $N$. In this case, one gets then (see Supplementary Material)
\begin{equation}\label{main3}
{\cal F}_{Q}(\hat\rho_0)\le\tilde C_Q\equiv\left[ \frac{2N}{1+\sqrt{1 + \frac{(1-\eta)N}{\eta}}} \right]^{2} \ .
\end{equation}
The behavior of the corresponding lower bound for the phase uncertainty $\delta\theta$, as a function of $N$, is shown in Fig.~\ref{fig3}, where the phase estimation error is normalized by the shot-noise limit, expressed in terms of the total number of photons in both arms \cite{Dornerprl09,Dempra09}: $\delta\theta_{SN}=(\sqrt{\eta}+1)/\sqrt{4\nu\eta N}$. This result clearly exhibits the change from the Heisenberg scale to a $1/\sqrt{N}$ scale. For $N>>\eta/(1-\eta)$, one gets a $1/\sqrt{N}$ scaling, corresponding to Eq.~(\ref{dtheta}), with the average number of photons replaced by $N$, while for  $N<<\eta/(1-\eta)$, the Heisenberg dependence, proportional to $1/N$, is obtained.

One may wonder how tight is the bound (\ref{main3}). The comparison of  this bound with  ${\cal F}_Q$ for the numerically determined optimal states is shown in Fig.~\ref{fig3}, for all $\eta$ and for different values of $N$ up to $N=100$. This yields, for this range of $N$,  $1/\sqrt{\nu\tilde C_Q}\le\delta\theta\le1.25/\sqrt{\nu\tilde C_Q}$, thus showing that, for these states, $\tilde C_Q$ provides a very good qualitative and quantitative approximation to the ultimate quantum limit.   As N increases, the fundamental limit of the phase uncertainty given by the quantum Fisher information becomes at most 1.25 larger than the one given by our bound.
 
\noindent {\bf Precision limits for atomic spectroscopy under dephasing}\\ 
We consider now the estimation of transition frequencies in atomic spectroscopy, in the presence of Markovian dephasing, which is a common source of decoherence for atoms, and the most important one in trapped-ions experiments. The aim is to estimate  the transition frequency $\omega_{0}$, by preparing the atoms in a known initial state, letting them evolve freely, and then measuring the final atomic state. The resources here, which affect the error in the estimation, are the number $N$ of atoms and the total time $T=t \nu$, where $t$ is the evolution time for each atom and $\nu$ is the number of experimental repetitions. In the absence of decoherence, the error scales as $\sim 1 / \sqrt{N}$ if the initial state of the $N$ atoms is separable. On the other hand, an initial maximally entangled GHZ atomic state may improve the scaling to $\sim 1 / N$ \cite{Bollingerpra96}. The situation changes when dephasing is present. In \cite{Ciracprl97}, it was shown that, in this case,  a separable state or a GHZ state of the atoms lead to the same error $\delta \omega_{0} = \sqrt{2 \gamma e / (N T)}$, where $\gamma$ is the dephasing rate.  It was also shown that a generalized Ramsey spectroscopy, involving a joint measurement of all the atoms, may lead, for an optimal non-maximally entangled state, to an error limited by $\delta \omega_{0} \geq \sqrt{2 \gamma / (NT)}$, a better bound then before, but still proportional to $1/\sqrt{N}$. In this regard, two questions were raised in Ref.~\cite{Ciracprl97}: whether this bound is saturated asymptotically by  the proposed measurement strategy; and whether it coincides with the one obtained from the quantum Fisher information.  Ref.~\cite{ulam}  has shown that, for properly chosen initial states,   this bound is actually saturated asymptotically, for the proposed measurement scheme.  Here we show that the remaining question can be tackled by our method. Indeed, we show that the above bound coincides asymptotically with the one given by the quantum Fisher information optimized over all initial atomic states. 

We show in the Supplementary Material that it is possible to choose a physically-motivated set of Kraus operators  such that our bound, optimized over all initial states, leads to
\begin{equation}
\nu{\cal F}_{Q}^{\rm max}\leq\dfrac{N T}{2 \gamma}\left[\dfrac{2\gamma t N}{1 +(e^{2\gamma t} - 1)N} \right]\leq\dfrac{N T}{2 \gamma} \ .
\label{boundat}
\end{equation}
which coincides with the bound derived in Ref.~\cite{Ciracprl97}.  Since Ref.~\cite{ulam} showed that this bound is attained asymptotically for the above generalized Ramsey measurement and a specific class of initial states, it follows that necessarily $\lim_{N\rightarrow\infty}\nu{\cal F}_{Q}^{\rm max}/N\ge T/2\gamma$. In view of inequality (\ref{boundat}), one concludes  that  $\lim_{N\rightarrow\infty}\nu{\cal F}_{Q}^{\rm max}/N = T/2\gamma$. This generalizes the previous results, showing that, no matter the initial state and the measurement scheme, the maximum improvement obtainable in the presence of dephasing is by a factor $\sqrt{e}$, which does not change the scaling of the error with $N$. Therefore, also in this case decoherence does not allow attaining the quantum scale, no matter how small the dephasing rate is, as long as $N$ is sufficiently large. It can be  shown that this conclusion remains valid in the presence of feedback.

\noindent {\bf Summary and perspectives}\\
We presented here a lower bound for the error in single-parameter estimation, within the framework of quantum metrology, valid for both unitary and non-unitary processes, which is always attainable.  The calculation of the best state leading to the ultimate quantum limit in the presence of noise is a difficult task, requiring numerical analysis that becomes more and more cumbersome as the number of resources increases. The upper bound (\ref{eqtot}), on the other hand, circumvents this difficulty, leading to useful relations that do not depend on the initial state, and that do not require optimizations over all possible Kraus representations. Indeed, convenient classes of Kraus operators may be chosen inspired by physical considerations regarding the process under investigation. The power of this method was exemplified within the framework of two important problems in quantum metrology, the estimation of phase in optical interferometry and of transition frequencies in atomic spectroscopy.  In optical interferometry, it captures the main features of the transition from the Heisenberg limit to the asymptotic shot-noise-like behavior. Furthermore, it imposes a severe restriction on the asymptotic behavior of the estimation error: even for weak noise, the improvement on the shot-noise limit is at most by a constant factor, which does not change the $1/\sqrt{N}$ behavior.  

Due to the omnipresence of non-unitary dynamics and the problem of parameter estimation in several fields of science, we envisage that this approach might find useful applications to other kinds of systems, involving for instance stochastic processes  \cite{andre} and dynamical evolutions that depend non-linearly on the number of resources. 
\\
\noindent {\bf Methods}
\\
\noindent {\bf Outline of the derivation of the upper bound to the quantum Fisher information}\\
Given an initial pure state $\hat{\rho}_{0}=\vert\psi\rangle\langle\psi\vert$ of a system $S$ and an arbitrary process, which changes the state to $\hat{\rho}(x)\equiv\sum_{\ell}\hat{\Pi}_{\ell}(x) \hat{\rho}_{0}\hat{\Pi}^{\dagger}_{\ell}(x)$, we expand the Hilbert space, introducing an environment $E$ so that the total state in $S+E$ undergoes a unitary evolution, described by 
\begin{equation}\label{Psi}
\vert\Psi(x)\rangle=\hat{U}_{S,E}(x)\vert\psi\rangle_{S}\vert 0\rangle_{E}=\sum_{\ell}\hat{\Pi}_{\ell}(x)\vert\psi\rangle_{S}\vert\ell\rangle_{E} \,.
\end{equation}
Here $\vert 0\rangle_{E}$ is the initial state of the environment. The states $\vert\ell\rangle_{E}$  form an orthogonal basis in $E$, which is independent of the parameter $x$ to be estimated and the Kraus operators $\hat{\Pi}_{\ell}(x)$ act on $S$. 

The upper bound for the quantum Fisher information is obtained from the inequality
\begin{equation}\label{inequality}
{\cal F}_{Q}\equiv\max_{\hat{E}_{j}^{(S)}\otimes\openone^{(E)}} F\left(\hat{E}_{j}^{(S)}\otimes\openone^{(E)}\right)\leq \max_{\hat{E}_{j}^{(S,E)}} F\left(\hat{E}_{j}^{(S,E)}\right)\equiv C_{Q}\,,
\end{equation}
where $F$ is the Fisher information defined in Eq.~(\ref{fisher}). This inequality results from the fact that, for ${\cal F}_{Q}$, the maximization is made for all POVMs contained in the $S$ space, while for $C_{Q}$ the maximization is for all POVMs in $S+E$. Therefore, $C_{Q}$ is an upper bound for ${\cal F}_{Q}$. The right-hand side of (\ref{inequality}) can be explicitly evaluated, since one is dealing in this case with a unitary evolution. It is then straightforward to obtain (\ref{eqtot}).

\noindent {\bf Saturation of the upper bound to the quantum Fisher information}\\
We show now that it is always possible to choose a set of Kraus operators such that the inequality in Eq. (\ref{inequality}) is transformed into an equality. The proof is based on Uhlmann's theorem \cite{uhlmann} and on the relation between Bures' fidelity $F_{B}(\hat\rho_1,\hat\rho_2)\equiv {\rm Tr}\sqrt{\hat\rho_1^{1/2}\hat\rho_2\hat\rho_1^{1/2}}$ and the quantum Fisher information \cite{Braunsprl94}. Uhlmann's theorem implies that \cite{Nielsenbook01}
\begin{equation}\label{uhlmann}
\left(F_{B}\left[\hat\rho(x_{\rm real}),\hat\rho(x)\right]\right)^2={\rm max}_{|\Psi(x)\rangle}|\langle\Phi(x_{\rm real})|\Psi(x)\rangle|^2\,,
\end{equation}
where $|\Phi(x_{\rm real})\rangle$ is an arbitrary purification of $\hat\rho(x_{\rm real})$ in $S+E$ and the maximization runs over all purifications $|\Psi(x)\rangle$ of $\hat\rho(x)$, also in $S+E$. On the other hand, Ref.~\cite{Braunsprl94} shows that, up to second order in $\delta x= x- x_{\rm real}$, 
\begin{equation}\label{caves}
\left(F_{B}\left[\hat\rho(x_{\rm real}),\hat\rho(x)\right]\right)^2=1-(\delta x/2)^2{\cal F}_Q\left[\hat\rho(x_{\rm real})\right]\,.
\end{equation}
Expanding the right-hand side of (\ref{uhlmann}) up to second order in $\delta x$, and comparing the resulting expression with the right-hand side of (\ref{caves}), one gets (see Supplementary Material) (\ref{fqcq}).Therefore $C_Q$ coincides with the quantum Fisher information for some choice of Kraus operators.  This proves the attainability of our bound. We show, in the supplementary material, that there is, in fact, an infinite number of Kraus representations that satisfy the above relation.

\noindent {\bf Upper bound for quantum Fisher information in lossy optical interferometry}\\
A convenient choice for the Kraus operators corresponding to an interferometer with losses in one of the arms is
\begin{equation}
\hat{\Pi}_\ell\left(\theta; \alpha\right) = \sqrt{\dfrac{\left(1-\eta\right)^{\ell}}{\ell!}}e^{i \theta \left( \hat{n} - \alpha \ell \right)} \eta^{\frac{\hat{n}}{2}} \hat{a}^{\ell} ,
\label{krausalpha}
\end{equation}
where $\hat{a}$ is the annihilation operator corresponding to the dispersive-arm mode, and $\eta$ quantifies the photon losses (from $\eta=1$, lossless case, to $\eta=0$, complete absorption). The parameter $\alpha$ defines a family of Kraus operators and it is used to minimize the value of $C_{Q}$. This choice of Kraus operators is inspired by physical considerations. For $\alpha=0$ ($\alpha=-1$) the process can be interpreted as a probabilistic photon absorption event, simulated by a beam splitter with transmissivity $\eta$ and after (before) it a phase shift $\theta$. If the beam splitter is placed after the dispersive element, then monitoring the environment leads to full recovery of the information, yielding therefore a non-reliable bound, which coincides with the one for the lossless case, while only partial information is obtained if the beam splitter is placed before the dispersive element (since in this case the deflected photons do not carry phase information). This last situation corresponds to the experimental setup used in \cite{kac}. It may lead to a bound better than the one corresponding to the first choice, although not as good as the one obtained by optimizing over all values of $\alpha$.

The Kraus operators given by Eq. (\ref{krausalpha})
lead to (see Supplementary Material)
\begin{eqnarray}\label{hh2}
\hat{H}_{1}\left(\theta; \alpha\right)\!\! &=&\!\!  \eta \left( 1-\eta \right) (1+\alpha)^{2} \hat{n} \nonumber\\
&+& \left[ 1 - \left( 1- \eta \right) (1+\alpha) \right]^{2} \hat{n}^2 , \\
\hat{H}_{2}\left(\theta; \alpha\right)\!\!  &=&\!\!  \left[ 1 - \left( 1- \eta \right) (1+\alpha) \right]\hat{n} \  .
\end{eqnarray}
 We get then,  for all real values of $\alpha$, 
\begin{eqnarray}
C_{Q}(\hat{\rho}_{0}; \alpha)&=& 4\left[ 1 - \left( 1- \eta \right) (1+\alpha) \right]^{2} \langle\Delta\hat{n}^2 \rangle_0 \nonumber \\ &+& 4 \eta \left( 1-\eta \right) (1+\alpha)^{2} \langle\hat{n} \rangle_0 \ ,
\label{cqmain}
\end{eqnarray}
where $\langle\hat n\rangle_0$ is the initial average number of photons in the dispersive arm, and $\langle\Delta\hat{n}^2 \rangle_0 $ is the corresponding variance. 

When $\alpha=-1$, one gets the trivial result ${\cal F}_{Q} \leq 4 \langle \Delta \hat{n}^2 \rangle_0$, while for $\alpha = 0$, ${\cal F}_{Q} \leq 4 \eta \left[ \eta \langle \Delta \hat{n}^2 \rangle_0 + \left( 1-\eta \right) \langle \hat{n} \rangle_0 \right]$. The factor $\eta$ before the brackets in this last expression comes from the energy loss, a classical effect. This expression implies that, only super-poissonian states ($\langle\Delta\hat{n}^2 \rangle_0 > \langle\hat{n} \rangle_0$) may lead to an improvement of the phase estimation over the standard quantum limit, which corresponds to ${\cal F}_{Q} = 4 \eta  \langle \hat{n}\rangle_0$.

For each value of $\langle\Delta\hat{n}^2 \rangle_0$ and $\langle\hat{n} \rangle_0$, the minimization of $C_{Q}$ with respect to $\alpha$ yields (\ref{main1}) -- see Supplementary  Material.

One should note that there is a choice of Kraus operators for which  $\hat H_2=0$ in (\ref{hh2}), and at the same time the term proportional to $\hat n^2$ in $\hat H_1$ vanishes: the one that corresponds to taking in (\ref{hh2}) $1+\alpha=1/(1-\eta)$. This implies, according to the discussion in the Supplementary Material,  that, even in the presence of feedback \cite{berryprl85,armemprl89,sandersprl010}, the scaling in the phase uncertainty cannot be better than the standard quantum limit. Even though Refs.~\cite{berryprl85,armemprl89,sandersprl010} deal with global estimation, our method still applies, since (\ref{deltaxfq}) is a lower bound for the uncertainty in global estimations, which is given by the Holevo variance \cite{Holbook82}.  
\\
\noindent {\bf Acknowledgments}
\\ 
The authors acknowledge financial support from the Brazilian funding agencies CNPq, CAPES and FAPERJ.  This work was performed as part of the Brazilian National Institute for Science and Technology on Quantum Information.
\\

\end{document}